\documentclass[reprint,english,superscriptaddress,preprintnumbers,amsmath,amssymb,aps,prc,nofootinbib]{revtex4-2}

\usepackage[utf8]{inputenc}  
\usepackage[T1]{fontenc}     

\usepackage{orcidlink}
\usepackage{physics}
\usepackage{graphicx}
\usepackage{dcolumn}
\usepackage{bm}
\usepackage{booktabs}
\usepackage{makecell}
\usepackage{siunitx}  

\usepackage{hyperref}

\usepackage[nameinlink]{cleveref}

\begin{document}

\title{Hadronic rescattering effects on net-proton cumulants from functional renormalization group calculations}

\author{Qianru Lin}
\affiliation{Key Laboratory of Quark and Lepton Physics (MOE) \& Institute of Particle Physics, Central China Normal University, Wuhan 430079, China}
\affiliation{Institute for Theoretical Physics, Goethe University, Max-von-Laue-Strasse 1, 60438 Frankfurt am Main, Germany}
\affiliation{Helmholtz Research Academy Hesse for FAIR (HFHF), GSI Helmholtz Center, Campus Frankfurt, Max-von-Laue-Straße 12, 60438 Frankfurt am Main, Germany}

\author{Shi Yin\,\orcidlink{0000-0001-5279-6926}}
\affiliation{Institute for Theoretical Physics, Justus Liebig University Gie\ss en, 35392 Gie\ss en, Germany}
\affiliation{Helmholtz Research Academy Hesse for FAIR (HFHF), Campus Gie\ss en, 35392 Gie\ss en, Germany}

\author{Jianing Li\,\orcidlink{0000-0001-7193-7237}}
\affiliation{Physikalisches Institut, Universit\"at Heidelberg, 69120 Heidelberg, Germany}
\affiliation{GSI Helmholtzzentrum für Schwerionenforschung, Planckstr. 1, 64291 Darmstadt, Germany}

\author{Hannah Elfner}
\affiliation{GSI Helmholtzzentrum für Schwerionenforschung, Planckstr. 1, 64291 Darmstadt, Germany}
\affiliation{Institute for Theoretical Physics, Goethe University, Max-von-Laue-Strasse 1, 60438 Frankfurt am Main, Germany}
\affiliation{Helmholtz Research Academy Hesse for FAIR (HFHF), GSI Helmholtz Center, Campus Frankfurt, Max-von-Laue-Straße 12, 60438 Frankfurt am Main, Germany}

\author{Fabian Rennecke\,\orcidlink{0000-0003-1448-677X}}
\affiliation{Institute for Theoretical Physics, Justus Liebig University Gie\ss en, 35392 Gie\ss en, Germany}
\affiliation{Helmholtz Research Academy Hesse for FAIR (HFHF), Campus Gie\ss en, 35392 Gie\ss en, Germany}

\author{Long-Gang Pang}
\affiliation{Key Laboratory of Quark and Lepton Physics (MOE) \& Institute of Particle Physics, Central China Normal University, Wuhan 430079, China}

\author{Jan M. Pawlowski\,\orcidlink{0000-0003-0003-7180}}
\affiliation{Institut f\"ur Theoretische Physik, Universit\"at Heidelberg, Philosophenweg 16, 69120 Heidelberg, Germany}
\affiliation{ExtreMe Matter Institute EMMI, GSI, Planckstra{\ss}e 1, D-64291 Darmstadt, Germany}

\date{\today}
\begin{abstract}

Net-proton cumulants in the Beam Energy Scan region of heavy-ion collisions are widely used to probe critical fluctuations associated with the conjectured critical endpoint of Quantum Chromodynamics (QCD). Most existing studies, however, concentrate on the initial-state or phase-transition contributions, while the impact of hadronic rescattering on these observables has not been fully quantified. To address this gap, we construct event-by-event proton and antiproton distributions from functional renormalization group (fRG) cumulants using the maximum entropy principle, and propagate the resulting particles through the hadronic transport model SMASH in a simplified spherical evolution setup. We systematically investigate how the hadronic cascade modifies net-proton cumulants at collision energies $\sqrt{s_{NN}}=3.0$, 3.9, 4.9, 7.2, and 7.7~GeV. In the canonical-ensemble framework, which enforces exact net-baryon number conservation, the higher-order cumulant signal---in particular the ratio $C_4/C_2$ at $\sqrt{s_{NN}}=4.9$~GeV---is strongly reduced during the early stage of the cascade; the suppression of $C_4/C_2$ reaches approximately $20\%$. The non-monotonic energy dependence inherited from the fRG input survives the hadronic evolution, but its magnitude is substantially modified. These results demonstrate that hadronic rescattering provides a non-negligible background effect that must be accounted for when extracting QCD critical-point signals from experimental data.
\end{abstract}

\maketitle

\section{Introduction}
The phase structure of QCD at finite temperature and baryon chemical potential is one of the most active research areas in nuclear and particle physics. Of particular interest is the possible existence of a critical endpoint (CEP) in the QCD phase diagram, where the chiral crossover separating the hadronic and quark-gluon plasma phases turns in to a second-order phase transition. Heavy-ion collisions (HICs) provide a unique experimental environment to explore this phase structure. Event-by-event fluctuations of conserved quantities, such as net-baryon number, net-electric charge, and net-strangeness, are regarded as sensitive probes of the QCD phase transition and the critical endpoint\cite{Stephanov:1998dy, Stephanov:1999zu, Asakawa:2000wh,Hatta:2003wn, Luo:2017faz}. 
Higher-order cumulants of net-proton distributions were recognized as especially promising probes, since they are directly connected to QCD susceptibilities in the grand canonical ensemble (GCE)\cite{Stephanov:2008qz, Athanasiou:2010kw, Bazavov:2020bjn, Borsanyi:2020fev, Fu:2023lcm}. 

On the theoretical side, cumulants in the phase structure have been extensively calculated within low energy effective models,  \cite{Skokov:2010uh, Almasi:2016zqf, 
Fu:2016tey, Fu:2015amv,Fu:2015naa, Fu:2018swz, Wen:2019ruz, Fu:2021oaw}, extrapolations from lattice QCD, \cite{Aoki:2006we,Borsanyi:2011sw,Bellwied:2015lba,Borsanyi:2018grb,Bazavov:2020bjn,Borsanyi:2020fev}, and related methods. More recently, they have been obtained from the equilibrium QCD phase structure computed in functional QCD, \cite{Fu:2019hdw, Gao:2020fbl, Gunkel:2021oya, Pawlowski:2025jpg, Fu:2026qnl, Wang:2026xwa}, see \cite{Isserstedt:2019pgx, Fu:2023lcm, Lu:2025cls, Lu:2026ezr, Zhao:2026mcp}. In the present work we shall specifically use the functional renormalisation group (fRG) results from \cite{Fu:2023lcm}. 

These calculations typically provide fluctuations at chemical freeze-out. However, hadronic rescattering and decays after chemical freeze-out can significantly modify the final-state observables. The theoretical interpretation of measured cumulants is further complicated by exact conservation laws (e.g., net-baryon number conservation), which suppress fluctuations especially in the intermediate-energy regime \cite{Bzdak:2012an,ALICE:2019nbs,Vovchenko:2021kxx,Fu:2023lcm,Li:2023kja}. Moreover, the finite lifetime and spatial extent of the fireball limit the growth of the correlation length near a critical point, weakening critical signatures during the dynamical evolution \cite{Berdnikov:1999ph,Stephanov:1999zu,Stephanov:2001zj,Tan:2025bsv}. More recent developments have incorporated the out-of-equilibrium evolution of momentum-dependent fluctuations into hydrodynamic descriptions near the critical point \cite{Kitazawa:2012at,Mukherjee:2015swa,Monnai:2016kud,Stephanov:2017ghc}. The time evolution of higher-order moments in the hadronic stage has also been discussed within diffusion–dissipation and transport frameworks \cite{Nahrgang:2011mg,Herold:2016uvv,Nahrgang:2018afz,Hammelmann:2023aza, Chen:2024zry}. The interplay between critical dynamics, fluctuation evolution, and experimental observables in heavy-ion collisions has been reviewed comprehensively in Ref.\ \cite{Bluhm:2020mpc}. Despite these efforts, a quantitative understanding of how hadronic cascade processes alter the equilibrium cumulants across different collision energies remains incomplete. 

Experimentally, the STAR Collaboration has performed extensive measurements of net-proton cumulants in the RHIC Beam Energy Scan program \cite{STAR:2014egu,Luo:2014tga,STAR:2021fge,STAR:2021iop,STAR:2022etb,STAR:2022vlo,STAR:2025zdq}, revealing a non-monotonic energy dependence of higher-order ratios that has been interpreted as a potential signature of critical dynamics \cite{STAR:2021rls,STAR:2022vlo,STAR:2025zdq}. 
Future measurements by the CBM experiment at FAIR \cite{Schmidt:2014lva} and recent results from the HADES Collaboration \cite{HADES:2020wpc,Nabroth:2025elh} extend these studies to even lower collision energies. Theoretical frameworks that quantitatively connect first-principles freeze-out calculations to these experimental observables are therefore essential. 

To quantify noncritical background effects from hadronic rescattering, initial-state fluctuations calculated via the fRG are coupled to the hadronic transport model SMASH \cite{Bass:1998ca,Bleicher:1999xi,Yan:2011fq,Voronyuk:2014rna,SMASH:2016zqf}. In particular, conservation effects in the SMASH framework were first explored in Ref.\cite{Hammelmann:2022yso}. Proton and antiproton distributions are reconstructed using the maximum entropy principle and propagated through SMASH to quantify the impact of the hadronic cascade. This framework enables a systematic investigation of net-proton cumulant evolution, focusing particularly on the energy dependence of the $C_4/C_2$ ratio across the Beam Energy Scan (BES) energies.

Following the procedure developed in Ref.~\cite{Hammelmann:2023aza} for an Ising-model parametrization of critical fluctuations, we explore here the hadronic rescattering effects on the equilibrium cumulants in  \cite{Fu:2023lcm}. The paper is structured as follows. Section~\ref{sec:background} describes our method for constructing proton and antiproton distribution functions, employing the maximum entropy principle with cumulants provided by fRG. Section~\ref{sec:results} presents the time evolution of these cumulants within the SMASH framework. Finally, Sec.~\ref{sec:Conclusion} summarizes the key findings and outlines their implications for future research.

\section{Method}
\label{sec:background}

Our approach consists of three main steps: (i) reconstructing the proton and antiproton multiplicity distributions from the equilibrium cumulants using the maximum entropy principle, (ii) generating initial-state particle configurations with kinematics sampled from thermal distributions, and (iii) evolving the system with the SMASH hadronic transport model.

To perform the dynamical evolution in SMASH, separate proton and antiproton distributions on an event-by-event basis are required. The fRG input from \cite{Fu:2023lcm} provides cumulants of the net-baryon number rather than separate proton and antiproton cumulants, so an additional modeling assumption is needed to construct the particle lists, $C_n^B \simeq C_n^P$. Additionally, in the collision-energy range considered here, $3.0 \le \sqrt{s_{NN}} \le 7.7$~GeV, the mean antiproton multiplicity is strongly suppressed, but it is not strictly zero on an event-by-event basis. Setting $N_{\bar p}=0$ in every event would therefore artificially remove residual antiproton-number fluctuations. Instead, we construct separate proton and antiproton multiplicities through an event-by-event matching procedure.

We first reconstruct the net-proton multiplicity distribution from the cumulants computed with fRG using the maximum entropy principle. Sampling from this distribution gives five million event-by-event net-proton numbers $N_p^{\rm net}=N_p-N_{\bar p}$. Independently, we sample a total proton multiplicity $N_p$ for each event. The antiproton number then follows from $N_{\bar p}=N_p-N_p^{\rm net}$, and the physical constraint $N_p^{\rm net}\le N_p$ is imposed to guarantee $N_{\bar p}\ge 0$. The accepted pairs $(N_p,N_{\bar p})$ provide the initial multiplicities fed into SMASH. The statistical uncertainties associated with finite sampling in cumulant calculations have been carefully studied in transport model simulations \cite{Steinheimer:2017dpb}.

\begin{figure}
\centering
\includegraphics[width=0.45\textwidth]{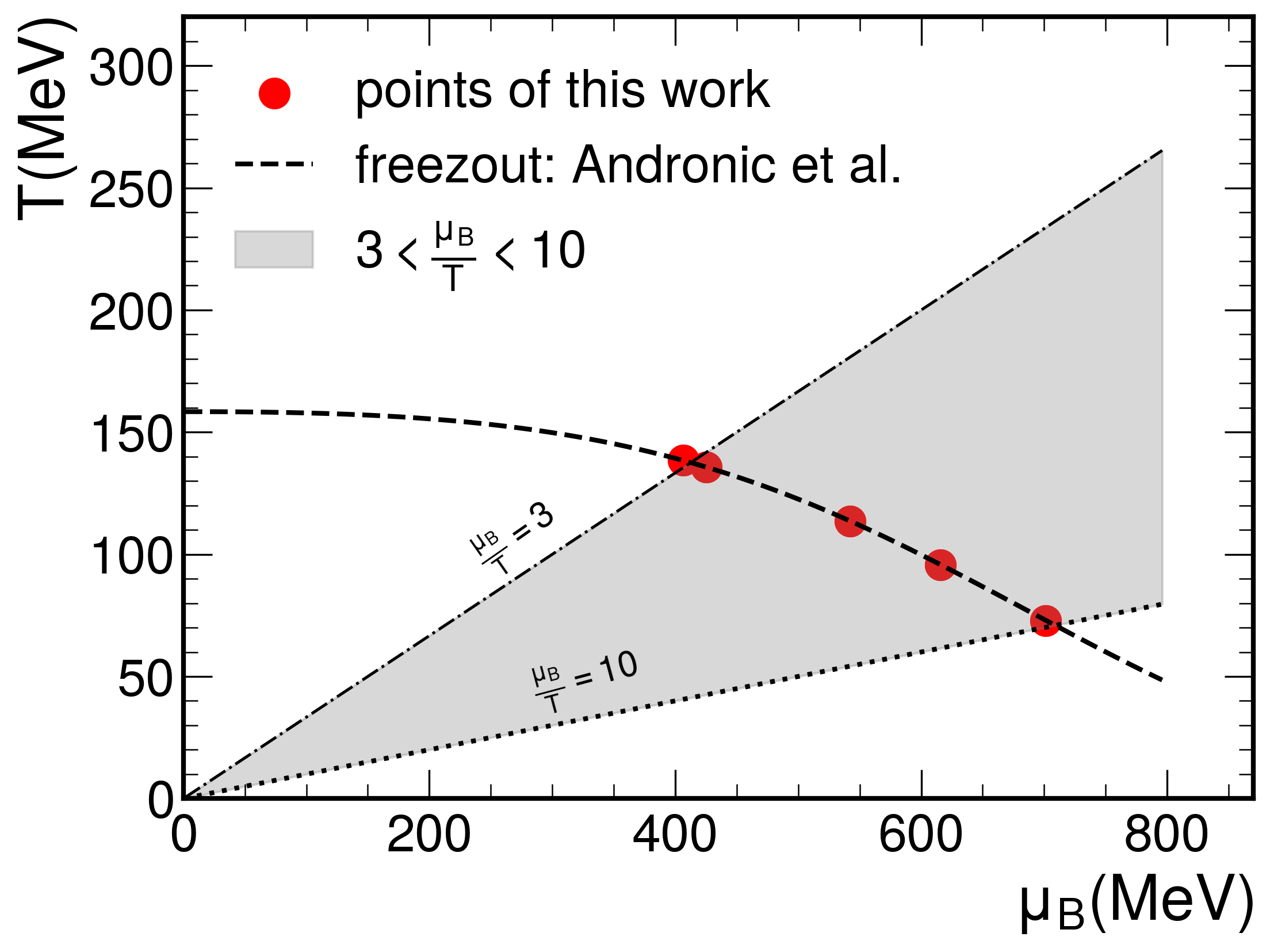}
\caption{Freeze-out temperature and baryon chemical potential as functions of collision energy. The curves are obtained from the parameterization of Andronic~\textit{et al.} \cite{Andronic:2017pug,Fu:2021oaw}.}
\label{fig:muB_T}
\end{figure}

For each pair $(N_p, N_{\bar{p}})$, the kinematic information of the particles is generated according to the corresponding freeze-out conditions: the temperature $T$ and baryon chemical potential $\mu_B$ taken from Figure.~\ref{fig:muB_T} at the given collision energy. The momentum magnitude is sampled from a relativistic thermal distribution $p^2 e^{-E/T}$, and the momentum direction is chosen isotropically. The particle four-momentum is then boosted by the radial flow velocity
\begin{equation}
\vec{u}(r) = \vec{e}_r \, \frac{u_0 r}{R},
\end{equation}
so that the momentum-space distribution at position $r$ is $f_{i,k}=\exp(-u \cdot k_i/T)$ \cite{Ling:2015yau,STAR:2017sal}. Here $k_i$ is the three-momentum of particle $i$, $T$ is the freeze-out temperature from the Andronic curve, and $u_0$ is the radial flow parameter taken from Table~\ref{tab:para_tabel}. The spatial coordinates are sampled uniformly within a sphere of radius $R=5.89$~fm, with a radial probability density proportional to $r^2$ and an isotropic angular distribution. In this way, we obtain a complete initial-state sample of protons and antiprotons that respects the higher-order cumulants from the fRG calculation. The spherical expansion with initial velocity mimics the final state evolution as long as full hybrid simulations including fluctuations are not yet existing.

Finally, the newly generated protons and antiprotons are passed as a particle list to the SMASH transport approach. The system is evolved up to $t=101$~fm/$c$ with the corresponding expansion velocity. By this time, the scattering rate has become negligible, indicating that kinetic freeze-out has been reached. 
The STAR transverse momentum cut $0.4<p_T<2$~GeV/$c$ and rapidity cut $|y|<0.5$ are not applied in the present setup, since the spherical expansion does not provide a physically meaningful rapidity variable; instead, the $p_T$ cut is replaced by a cut on the total momentum, $0.4<p<2$~GeV/$c$ with $p=\sqrt{p_x^2+p_y^2+p_z^2}$. The cumulants of the net-proton distribution are then recalculated, allowing us to investigate the influence of hadronic rescattering on the final-state observables.

\subsection{From fRG susceptibilities to cumulants}
\label{sec:fRGcumulants}

Functional QCD computations in equilibrium QCD provide generalized baryon-number susceptibilities in the grand canonical ensemble (GCE),
\begin{equation}
\chi_n^B=\frac{\partial^n}{\partial(\mu_B/T)^n}\frac{p}{T^4},
\end{equation}
from which the GCE baryon-number cumulants are obtained as
\begin{equation}
C_n^{B,\mathrm{GCE}}=(VT^3)\,\chi_n^B\,,
\end{equation}
with $V$ for the total system volume. Since ratios such as $C_4/C_2$ are independent of the overall system volume, we work directly with these volume-scaled cumulants. We use the fRG results of Ref.~\cite{Fu:2023lcm}, that are based on the QCD phase structure and QCD dynamics from \cite{Fu:2019hdw} (fRG). 

These results provide cumulants of the net-baryon number, whereas the experimentally measured observable is the net-proton number. Mapping baryon-number cumulants onto proton-number cumulants is intrinsically model dependent because it requires the identification of hadronic quantum numbers at the freeze-out stage. In the following we adopt the standard approximation $C_n^B \simeq C_n^P$ \cite{Vovchenko:2017gkg,Vovchenko:2020tsr,Braun-Munzinger:2020jbk}, i.e. we identify net-baryon cumulants with net-proton cumulants. This identification is expected to be reasonable at high collision energies, where the proton fraction among baryons is large and the GCE is a good approximation, but it becomes more uncertain at the lower energies studied here. 

For a finite heavy-ion system, exact net-baryon number conservation suppresses fluctuations relative to the GCE. We account for this by applying the subensemble acceptance method (SAM) \cite{Vovchenko:2020tsr}, which converts the GCE susceptibilities into cumulants in the canonical ensemble (CE). In the SAM, an acceptance fraction $\alpha=V_1/V$ describes the ratio of the detector acceptance volume $V_1$ to $V$, and the CE cumulants up to fourth order follow from the relations given in Eqs.~(9)--(13) of Ref.\ \cite{Vovchenko:2020tsr}. The value of $\alpha$ is fixed phenomenologically by matching to the measured proton distributions before rescattering. In this work, we employ $\alpha = 0.25$, $0.22$, $0.19$, $0.13$, and $0.12$ at $\sqrt{s_{NN}} = 3.0$, $3.9$, $4.9$, $7.2$, and $7.7$~GeV, respectively~\cite{Fu:2023lcm}.

The net-proton cumulants are defined in the usual way,
\begin{equation}
\begin{split}
C_1&=\langle N_p^{\rm net}\rangle,\\
C_2&=\langle (\delta N_p^{\rm net})^2\rangle,\\
C_3&=\langle (\delta N_p^{\rm net})^3\rangle,\\
C_4&=\langle (\delta N_p^{\rm net})^4\rangle-3\langle (\delta N_p^{\rm net})^2\rangle^2,
\end{split}
\end{equation}
where $\langle\cdots\rangle$ denotes the ensemble average and $\delta N_p^{\rm net}\equiv N_p^{\rm net}-\langle N_p^{\rm net}\rangle$ is the event-by-event fluctuation of the net-proton number. Both the GCE cumulants $C_n^{B,\mathrm{GCE}}$ and the CE cumulants obtained through the SAM are used as input to the maximum-entropy reconstruction described below, providing a benchmark for the role of exact conservation laws.

\subsection{Maximum entropy method}
\label{sec:MEM}

To reconstruct the probability distribution function, knowledge of all cumulants, or equivalently all moments, is in principle required. In this work we use input for the first four cumulants from \cite{Fu:2023lcm}. We therefore employ the maximum entropy principle to reconstruct the probability distribution with the least additional assumptions. In the absence of further information, the maximum entropy distribution is the least biased distribution consistent with the known constraints~\cite{10.1063/1.526446}.

The input from \cite{Fu:2023lcm} provides the first four cumulants ($C_1,\ldots,C_4$). Before applying the maximum entropy method, these cumulants are converted into the corresponding raw moments via the general recursions
\begin{equation}
\mu_n = \sum_{m=0}^{n-1} \binom{n-1}{m} C_{n-m}\,\mu_m\,,\quad{n=1,\,2,\,3\cdots}
\end{equation}
The raw moments of $N_p^{\rm net}$ are defined as
\begin{equation}
\mu_k = \langle (N_p^{\rm net})^k\rangle, \qquad k=1,\ldots,4.
\end{equation}
The probability distribution $P(N_p^{\rm net})$ is then obtained by maximizing the Shannon entropy,
\begin{equation}
S = -\sum_{N_p^{\rm net}=-\infty}^{+\infty} P(N_p^{\rm net})\ln P(N_p^{\rm net}),
\end{equation}
subject to the normalization condition and the first four moment constraints,
\begin{equation}
\label{eq:constr}
\begin{split}
\sum_{N_p^{\rm net}=-\infty}^{+\infty} P(N_p^{\rm net}) &= 1,\\\sum_{N_p^{\rm net}=-\infty}^{+\infty}(N_p^{\rm net})^k P(N_p^{\rm net}) &= \mu_k,\qquad k=1,\ldots,4.
\end{split}
\end{equation}
Introducing Lagrange multipliers for the normalization and moment constraints, the maximum-entropy solution can be written as
\begin{equation}
P(N_p^{\rm net}) = \frac{1}{Z}\exp\left[-\sum_{k=1}^{4}\lambda_k (N_p^{\rm net})^k\right],
\end{equation}
where $Z$ is the normalization factor and $\lambda_k$ are the Lagrange multipliers associated with the moment constraints. They are determined by requiring that the reconstructed distribution reproduces the prescribed moments, see Eq.\ \ref{eq:constr}.
In practice, the resulting nonlinear equations for $\lambda_k$ are solved numerically. The obtained distribution $P(N_p^{\rm net})$ is then used to sample the initial event-by-event net-proton multiplicities for the subsequent SMASH evolution.

\subsection{Simulation parameters}
The hadronic cascade simulations are performed with SMASH version~3.2. The system is initialized in a spherical volume with radius $R=5.89$~fm. We generate five million events per collision energy, and the hadronic evolution is followed up to $t=101$~fm/$c$. The chemical freeze-out temperature $T$, baryon chemical potential $\mu_B$, and average transverse radial flow velocity $u_0$ for each collision energy are summarized in Table~\ref{tab:para_tabel} \cite{Nonaka:2004pg,STAR:2017sal,Andronic:2017pug,Fu:2021oaw}. The parameterizations used for $T(\sqrt{s_{NN}})$ and $\mu_B(\sqrt{s_{NN}})$ are taken from Ref.\ \cite{Andronic:2017pug}.

\begin{table}[htbp]
\centering
\caption{Chemical freeze-out parameters and average radial flow velocity for the collision energies studied in this work.}
\label{tab:para_tabel}
\begin{tabular}{cccc}
\hline\hline
$\sqrt{s_{NN}}$~(GeV) & $T$~(GeV) & $\mu_{B}$~(GeV) & $u_0$ \\
\hline
3.0 & 0.0729 & 0.7015 & 0.30 \\
3.9 & 0.0958 & 0.6158 & 0.35 \\
4.9 & 0.1136 & 0.5423 & 0.40 \\
7.2 & 0.1357 & 0.4254 & 0.43 \\
7.7 & 0.1384 & 0.4064 & 0.46 \\
\hline\hline
\end{tabular}
\end{table}

\subsection{Rescattering and decays}

To quantify the collision dynamics that drives the cumulant evolution, we record, for each time interval $\Delta t$, the number of protons that participate in collisions of a given SMASH process type, normalized to the total proton number at the beginning of the interval and to $\Delta t$. This yields the average collision rate per proton per unit time as a function of time for elastic scattering, resonance formation, and inelastic binary scattering. The collision-rate analysis is carried out on a subset of 12,500 independent events; the much larger sample of five million events is used for the statistical evaluation of the cumulants and their ratios.

\section{Results}
\label{sec:results}

In this section, we present the effects of hadronic rescattering on the net-proton cumulants obtained from the fRG calculations. We discuss the time evolution of the cumulant ratio $C_4/C_2$ in both the GCE and CE frameworks, followed by an analysis of the underlying collision dynamics.

As discussed in Sec.~\ref{sec:background}, the CE framework enforces exact net-baryon number conservation and is therefore better suited to the low collision energies considered here. In the following we present the CE results as the primary physical predictions of this work, while the GCE results serve as a benchmark for comparison with previous freeze-out calculations.

\subsection{Grand canonical benchmark}

\begin{figure}
\centering
\includegraphics[width=0.45\textwidth]{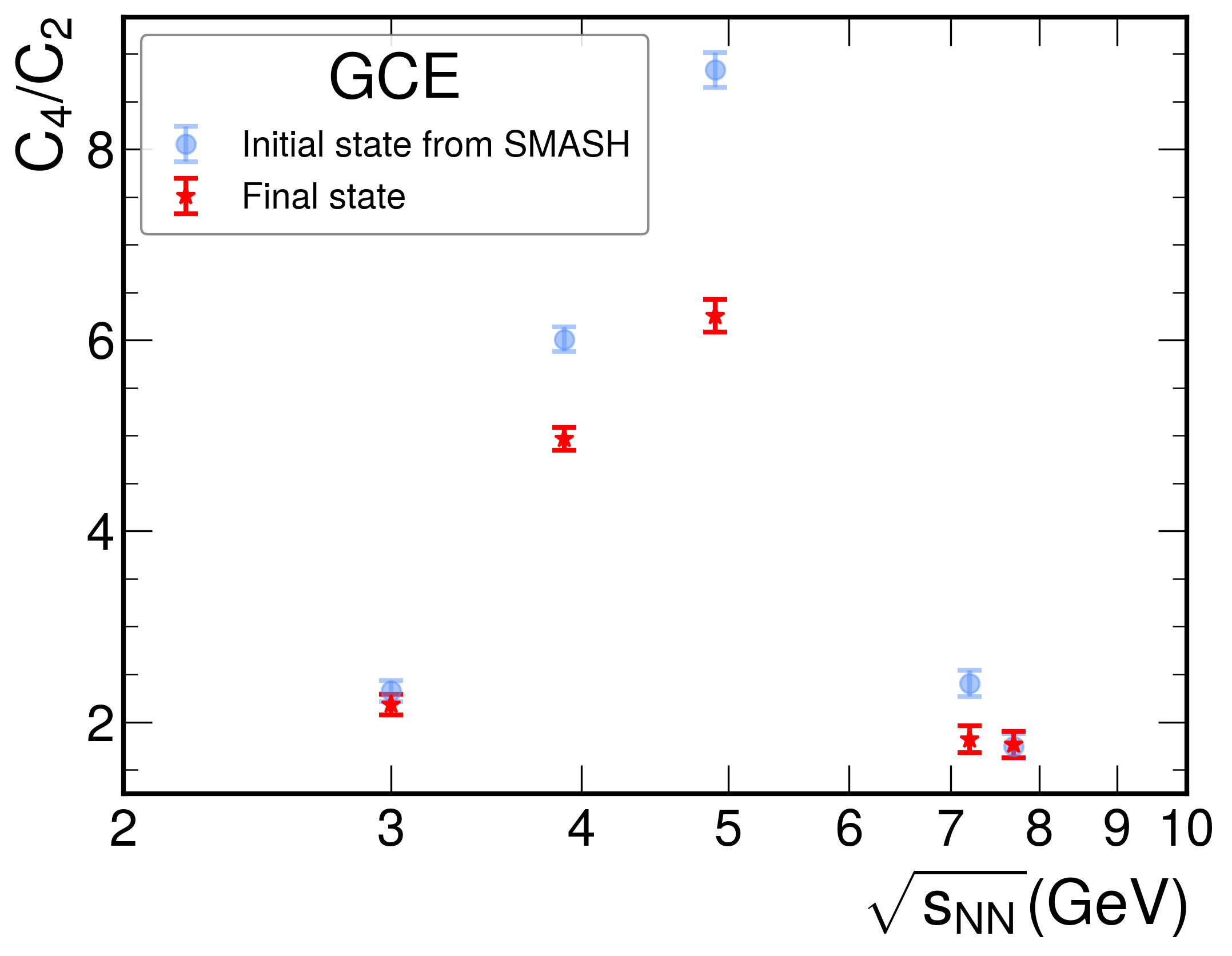}
\caption{Collision-energy dependence of the net-proton cumulant ratio $C_4/C_2$ in the GCE before (blue circles) and after (red stars) the SMASH evolution.}
\label{fig:GCE_C4C2_energy}
\end{figure}

Figure~\ref{fig:GCE_C4C2_energy} shows the initial $C_4/C_2$ ratio and the result after rescattering as functions of collision energy. Near 4.9~GeV the initial value displays a pronounced peak, while the final-state value is clearly suppressed. Although the magnitude changes during the evolution, the non-monotonic energy dependence remains visible, indicating that rescattering does not fundamentally alter the qualitative energy dependence of this observable.

\begin{figure}
\centering
\includegraphics[width=0.45\textwidth]{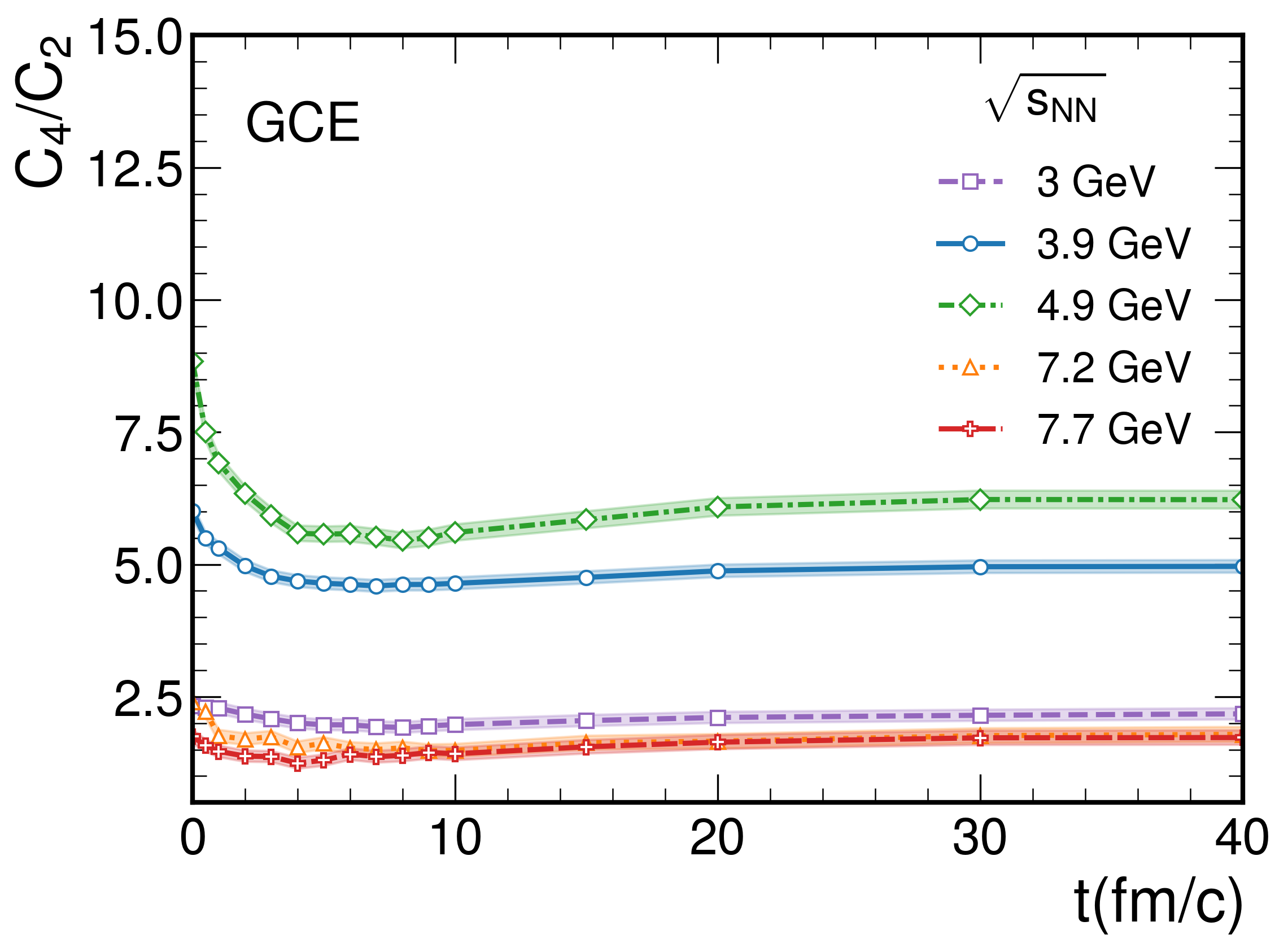}
\caption{Time evolution of the cumulant ratio $C_4/C_2$ of net protons in the GCE for the five collision energies considered. The shaded bands represent statistical uncertainties.}
\label{fig:GCE_C4C2_time}
\end{figure}

The GCE calculation is presented as a reference to isolate the effects of exact baryon-number conservation. It is not regarded as the primary physical description of the intermediate-energy systems considered in this work.

Figure~\ref{fig:GCE_C4C2_time} presents the time evolution of the cumulant ratio $C_4/C_2$ in the GCE. The shaded bands reflect statistical uncertainties estimated from the standard errors of the sample central moments. Despite the large event statistics of five million, the error bands remain sizable at certain energies, because the variance of the fourth-order cumulant estimator involves the eighth-order moment and is therefore strongly amplified for distributions with heavy tails.
Notably, the ratio exhibits a clear energy dependence: at $\sqrt{s_{NN}}=4.9$~GeV, $C_4/C_2$ shows the most pronounced suppression throughout the evolution, especially at late times. Since the fRG input already contains fluctuations associated with the assumed CEP, this pronounced suppression indicates that hadronic rescattering strongly modifies these pre-existing signals, rather than providing independent evidence for the CEP location.

\begin{figure}
\centering
\includegraphics[width=0.45\textwidth]{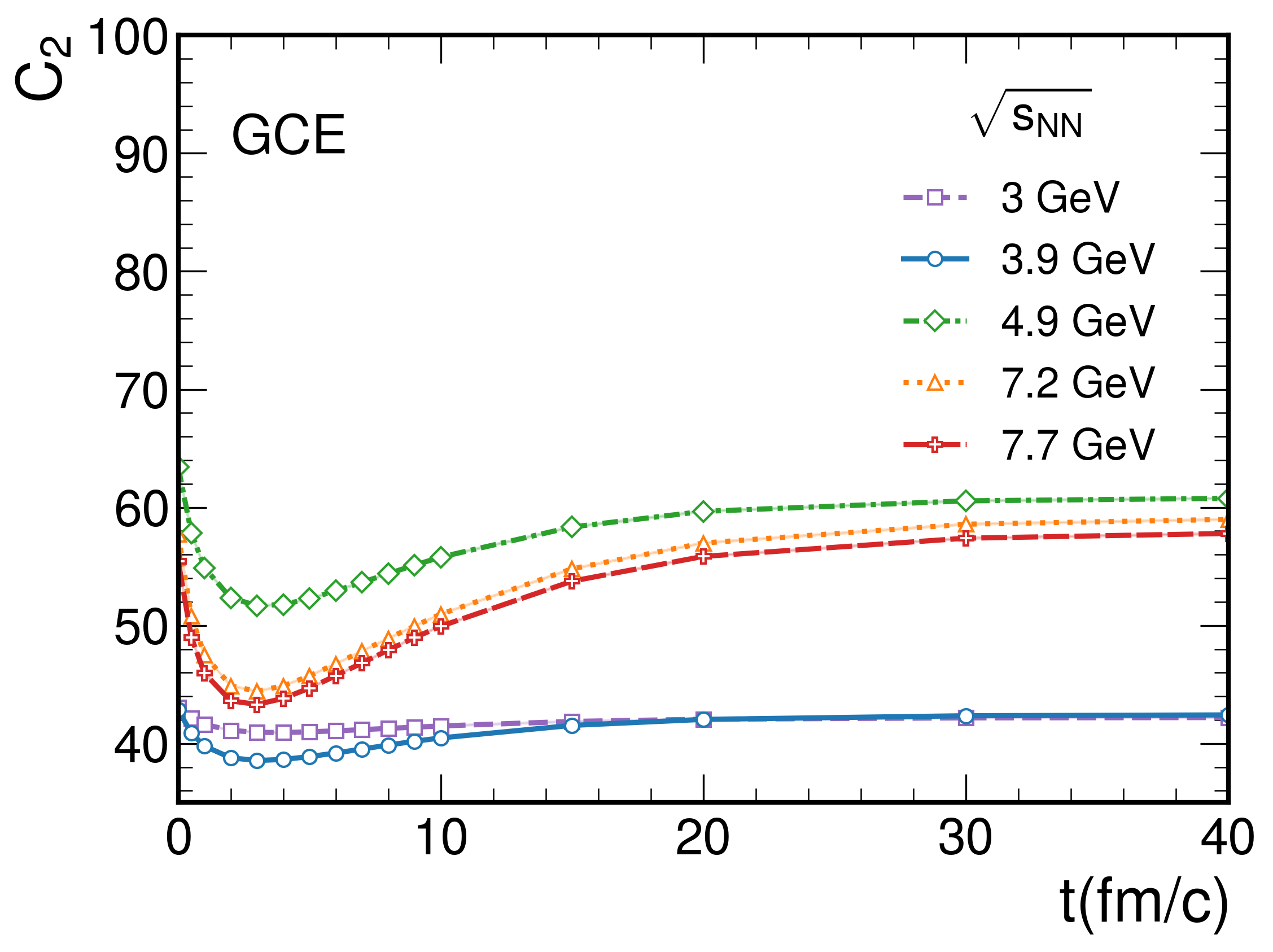}
\caption{Time evolution of the second-order cumulant $C_2$ of net protons in the GCE for the five collision energies considered. The shaded bands represent statistical uncertainties.}
\label{fig:GCE_C2_time}
\end{figure}

Figure~\ref{fig:GCE_C2_time} shows the time evolution of the $C_2$ in the GCE for the five collision energies considered in this work. At all energies, $C_2$ undergoes a noticeable modification during the early stage of the hadronic cascade. The values initially decrease, reach a minimum within the first several fm/$c$, and subsequently increase toward approximately stationary late-time values. This non-monotonic temporal behavior indicates that the hadronic evolution does not lead to a simple monotonic damping of the fluctuations. Instead, the early interactions reduce the variance of the net-proton multiplicity distribution, while the later evolution partially restores it as the system expands and the reaction rates decrease.

\begin{figure}
\centering
\includegraphics[width=0.45\textwidth]{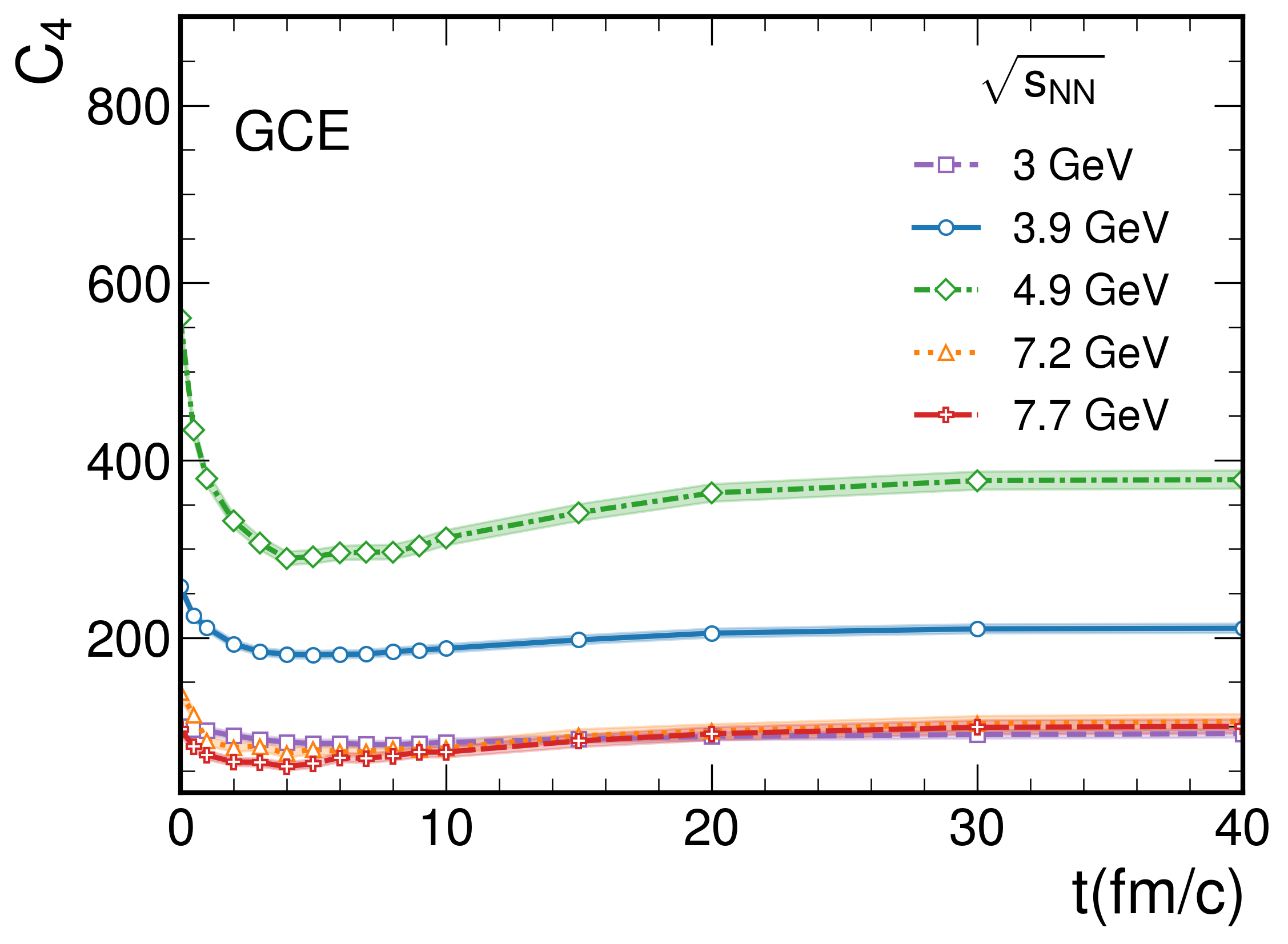}
\caption{Time evolution of the fourth-order cumulant $C_4$ of net protons in the GCE for the five collision energies considered. The shaded bands represent statistical uncertainties.}
\label{fig:GCE_C4_time}
\end{figure}

The magnitude of the early-time reduction depends on the collision energy. The decrease is relatively weak at 3.0~GeV, whereas a more pronounced suppression is visible at the higher collision energies, particularly at 7.2 and 7.7~GeV. At $\sqrt{s_{NN}}=4.9$ GeV, $C_2$ also decreases during the early evolution, but its relative reduction is smaller than the corresponding reduction of $C_4$ shown in Figure.~\ref{fig:GCE_C4_time}. This difference between the evolution of the second- and fourth-order cumulants is essential for understanding the behavior of the ratio $C_4/C_2$.

Figure~\ref{fig:GCE_C4_time} shows the time evolution of the $C_4$ in the GCE at different collision energies. At all energies, $C_4$ is modified predominantly during the early stage of the hadronic cascade. It initially decreases, reaches a minimum within the first several fm/$c$, and subsequently exhibits a partial recovery before approaching an approximately stationary value at late times. This behavior indicates that the hadronic evolution does not simply erase the initial fourth-order fluctuations through monotonic damping. Instead, the cumulant is strongly modified during the dense interaction stage and is partially restored during the subsequent expansion.

The most pronounced early-time reduction is observed at $\sqrt{s_{NN}}=4.9$~GeV. At this energy, $C_4$ starts from the largest initial value among the collision energies considered and decreases more rapidly and by a larger relative amount than the neighboring-energy results. Although it partially recovers at later times, its final value remains below the initial one. The results at the other collision energies also show an early decrease followed by a recovery, but the relative modifications are generally less pronounced. The particularly strong response at 4.9~GeV suggests that the enhanced fourth-order fluctuations inherited from the fRG input are more strongly affected by the hadronic evolution.

\subsection{Canonical ensemble results}

\begin{figure}
\centering
\includegraphics[width=0.45\textwidth]{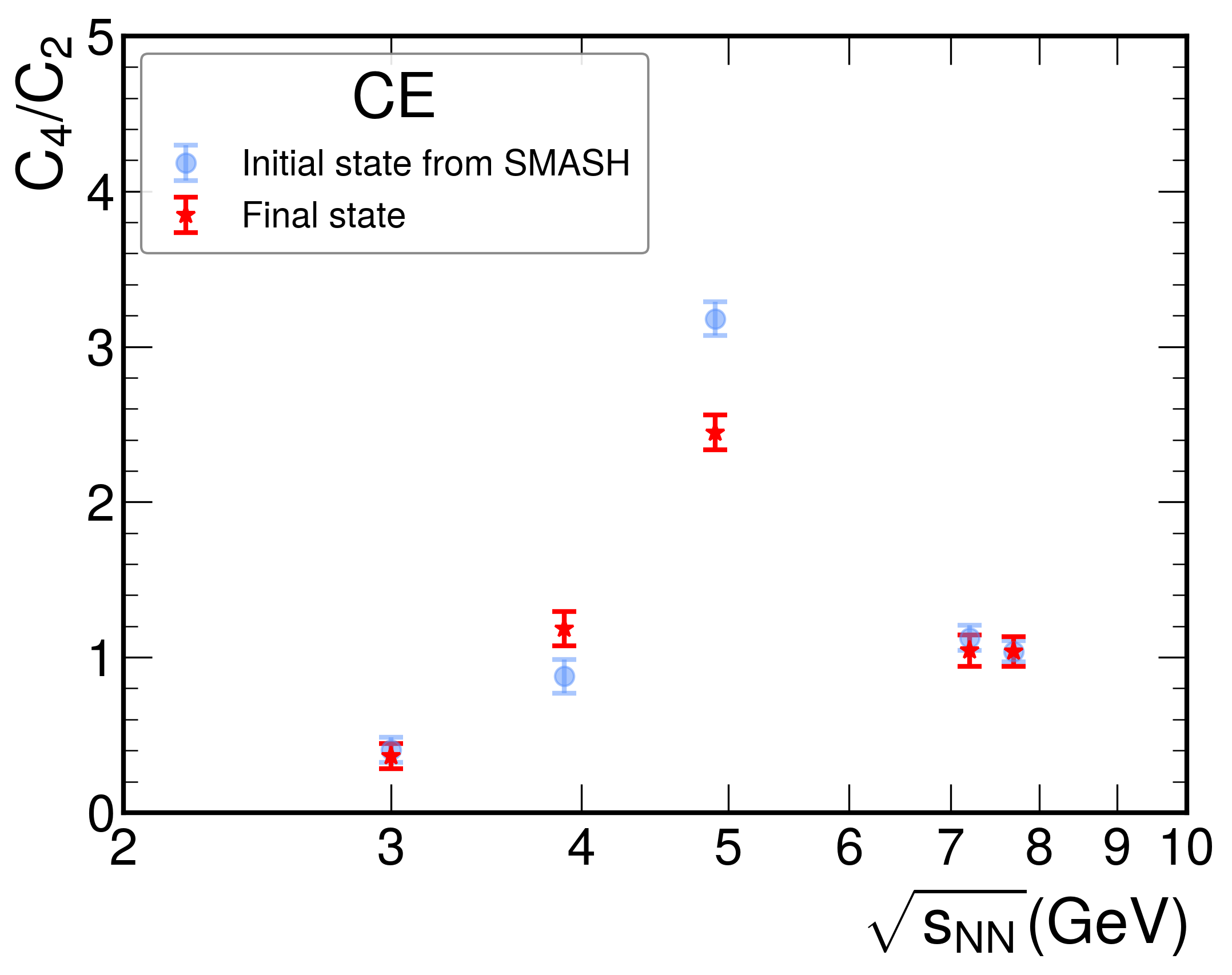}
\caption{Collision-energy dependence of the net-proton cumulant ratio $C_4/C_2$ in the CE before (blue circles) and after (red stars) the SMASH evolution.}
\label{fig:CE_C4C2_energy}
\end{figure}

Figure~\ref{fig:CE_C4C2_energy} summarizes the collision-energy dependence of $C_4/C_2$ in the CE before and after the hadronic cascade. The initial distribution exhibits a pronounced non-monotonic structure around $\sqrt{s_{NN}}=4.9$~GeV. After the SMASH evolution, the value at 4.9~GeV is reduced more strongly than those at the neighboring energies. Nevertheless, the non-monotonic structure remains visible in the final state.

\begin{figure}
\centering
\includegraphics[width=0.45\textwidth]{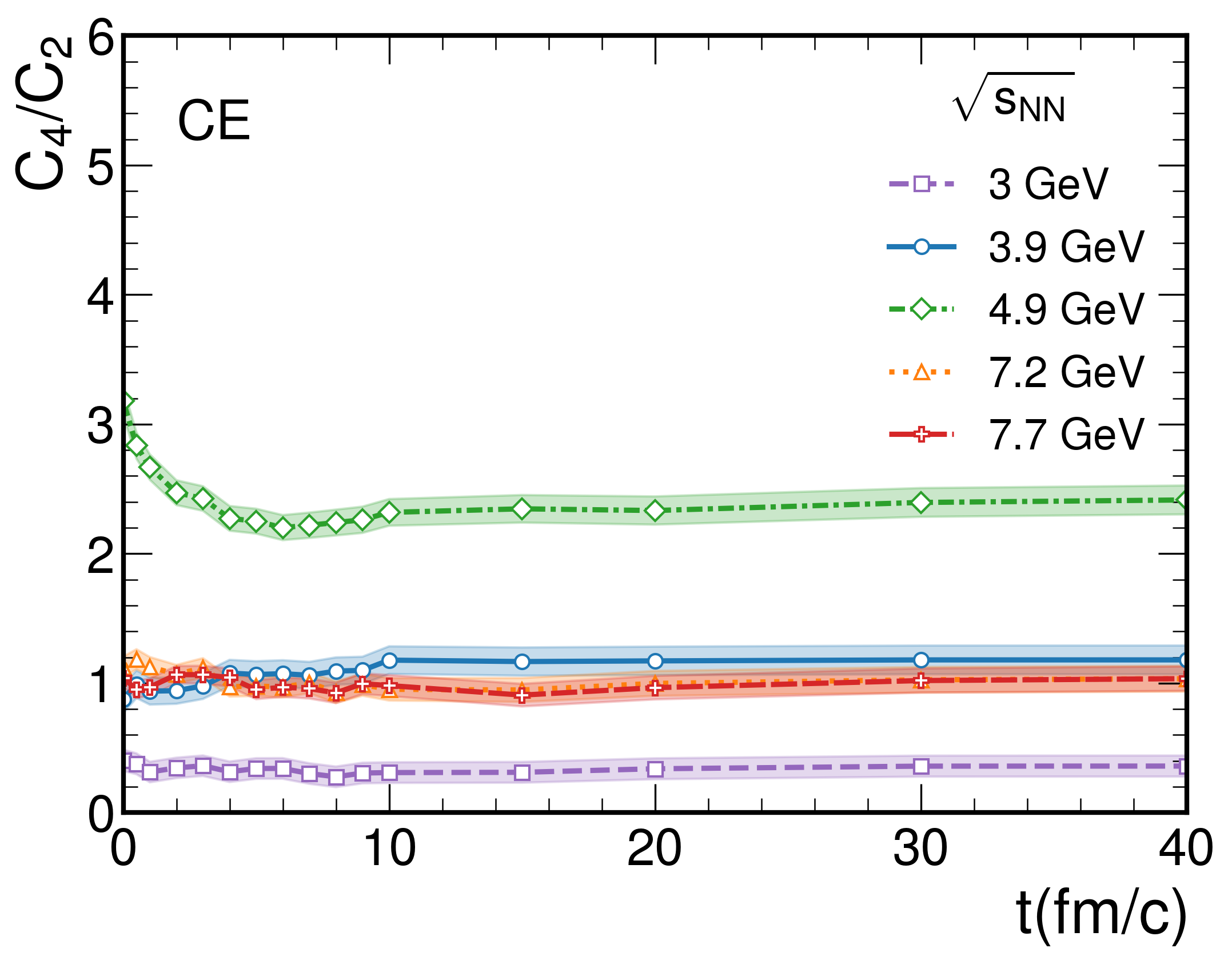}
\caption{Time evolution of the net-proton cumulant ratio $C_4/C_2$ in the CE for the five collision energies considered. The shaded bands represent statistical uncertainties.}
\label{fig:CE_C4C2_time}
\end{figure}

Figure~\ref{fig:CE_C4C2_time} shows the time evolution of $C_4/C_2$ in the CE. Among the collision energies considered, the result at $\sqrt{s_{NN}}=4.9$~GeV exhibits the largest relative reduction with respect to its initial value. The ratio decreases rapidly during the early stage of the hadronic evolution, reaches a minimum, and subsequently shows a partial recovery as the system expands and becomes dilute. We emphasize that the absolute value of $C_4/C_2$ at 4.9~GeV remains larger than those at the neighboring collision energies. Its distinctive feature is therefore the stronger relative suppression during the hadronic stage, rather than a smaller absolute magnitude.

This result indicates that hadronic rescattering quantitatively weakens the peak structure inherited from the fRG input, but does not completely erase its collision-energy dependence. The CE results therefore suggest that a critical-like non-monotonic structure may survive the hadronic stage, although its magnitude can be substantially modified before experimental detection.

\begin{figure}
\centering
\includegraphics[width=0.48\textwidth]{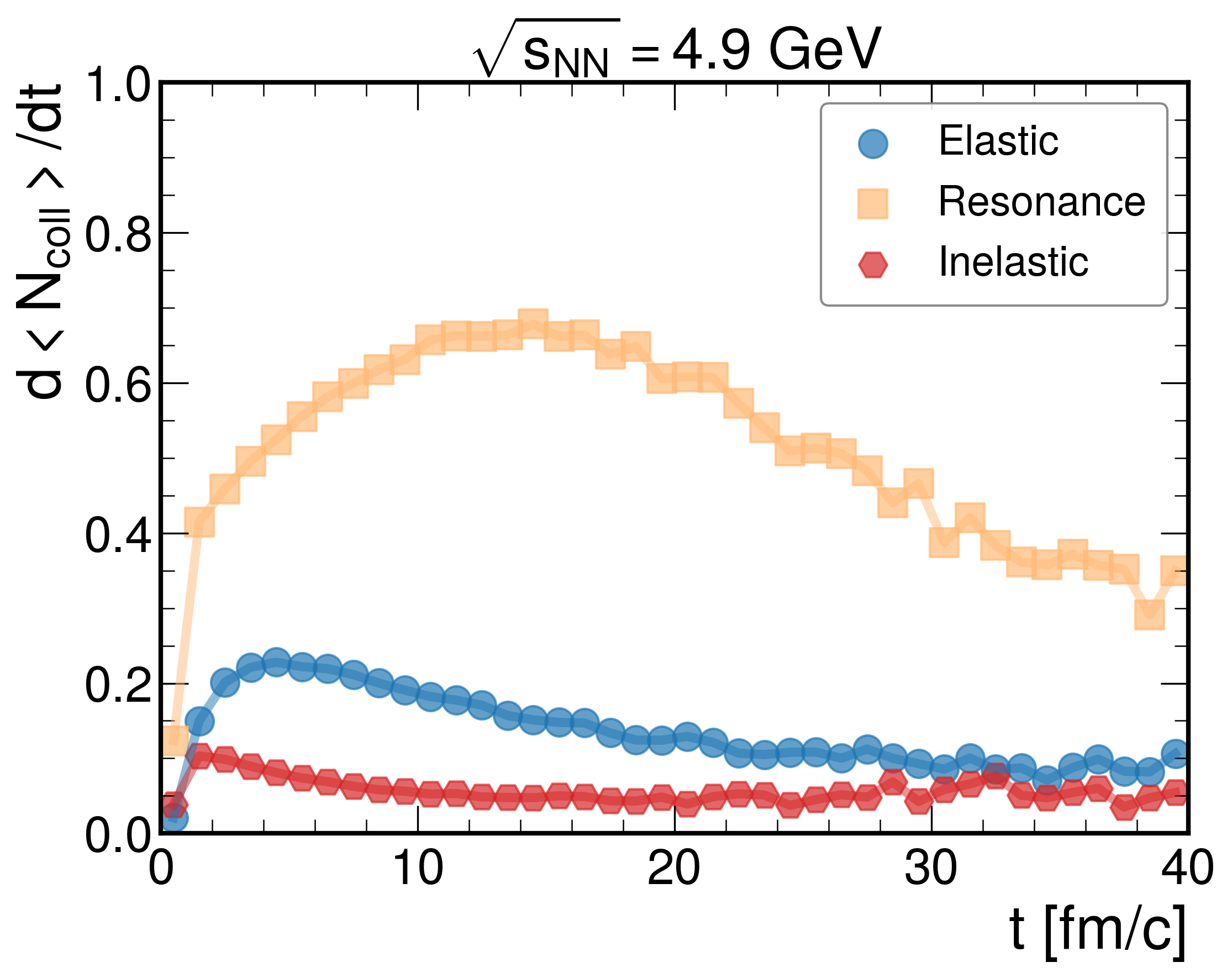}
\caption{Time evolution of the proton collision rate per proton for elastic scattering, resonance formation, and inelastic binary scattering at $\sqrt{s_{NN}}=4.9$~GeV.}
\label{fig:dNcoll_dt_4p9}
\end{figure}

\begin{figure}
\centering
\includegraphics[width=0.48\textwidth]{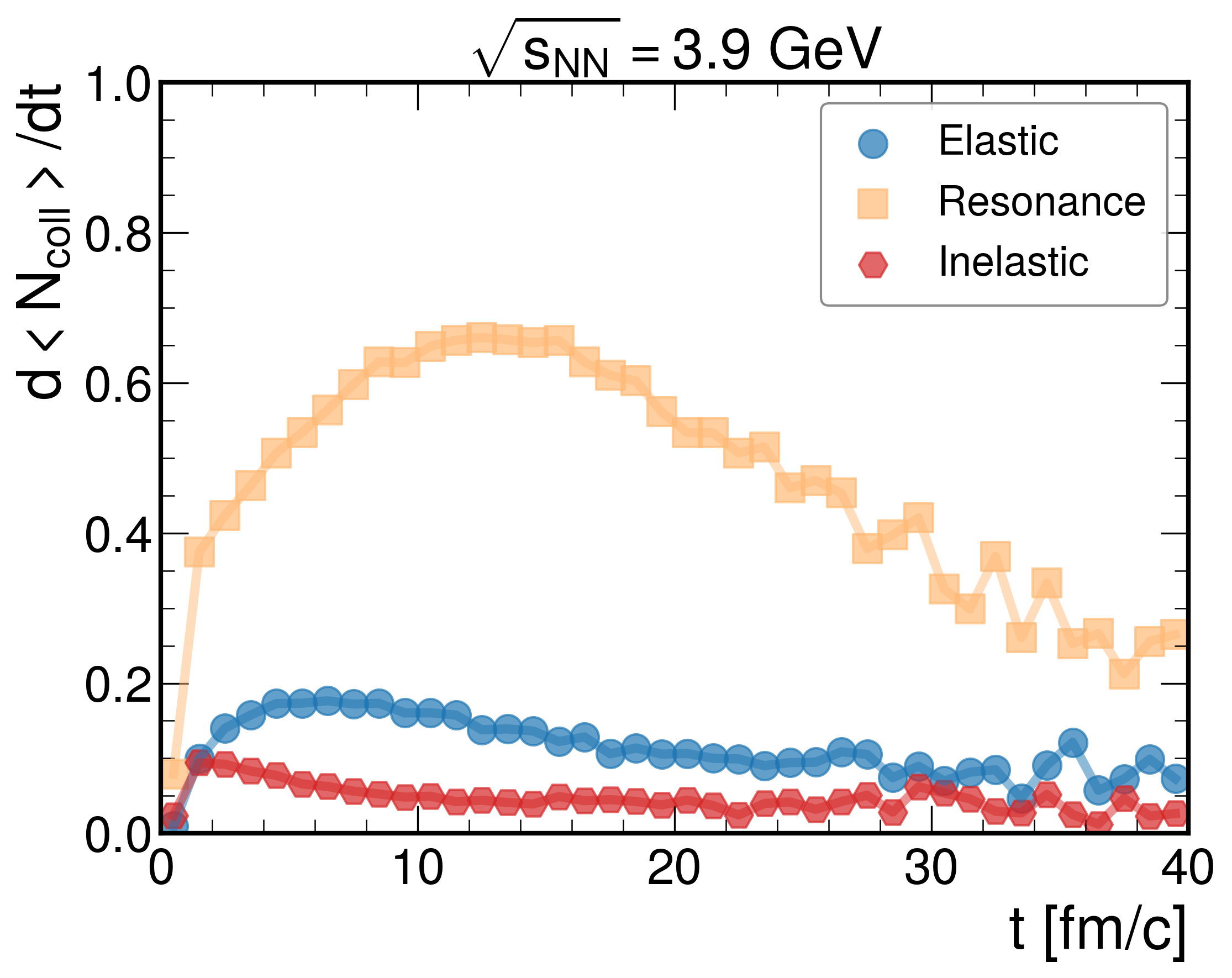}
\caption{Time evolution of the proton collision rate per proton for elastic scattering, resonance formation, and inelastic binary scattering at $\sqrt{s_{NN}}=3.9$~GeV.}
\label{fig:dNcoll_dt_3p9}
\end{figure}

Figure~\ref{fig:dNcoll_dt_4p9} and \ref{fig:dNcoll_dt_3p9} show the time evolution of the average proton collision rate per proton for the dominant scattering channels at $\sqrt{s_{NN}}=4.9$ and $3.9$ GeV, respectively. In both cases, the collision rate is largest during the early stage of the hadronic evolution and decreases as the system expands and becomes dilute. Resonance formation is the dominant process, followed by elastic scattering, while inelastic binary scattering remains subdominant. Compared with the $4.9$ GeV case, the overall collision rate at $3.9$ GeV is lower, indicating a weaker hadronic interaction activity at the lower collision energy. These early-stage interactions redistribute the proton-number fluctuations and constitute the main microscopic origin of the cumulant modifications observed in Figures. \ref{fig:GCE_C4_time}--\ref{fig:CE_C4C2_energy}.

\begin{figure}
\centering
\includegraphics[width=0.48\textwidth]{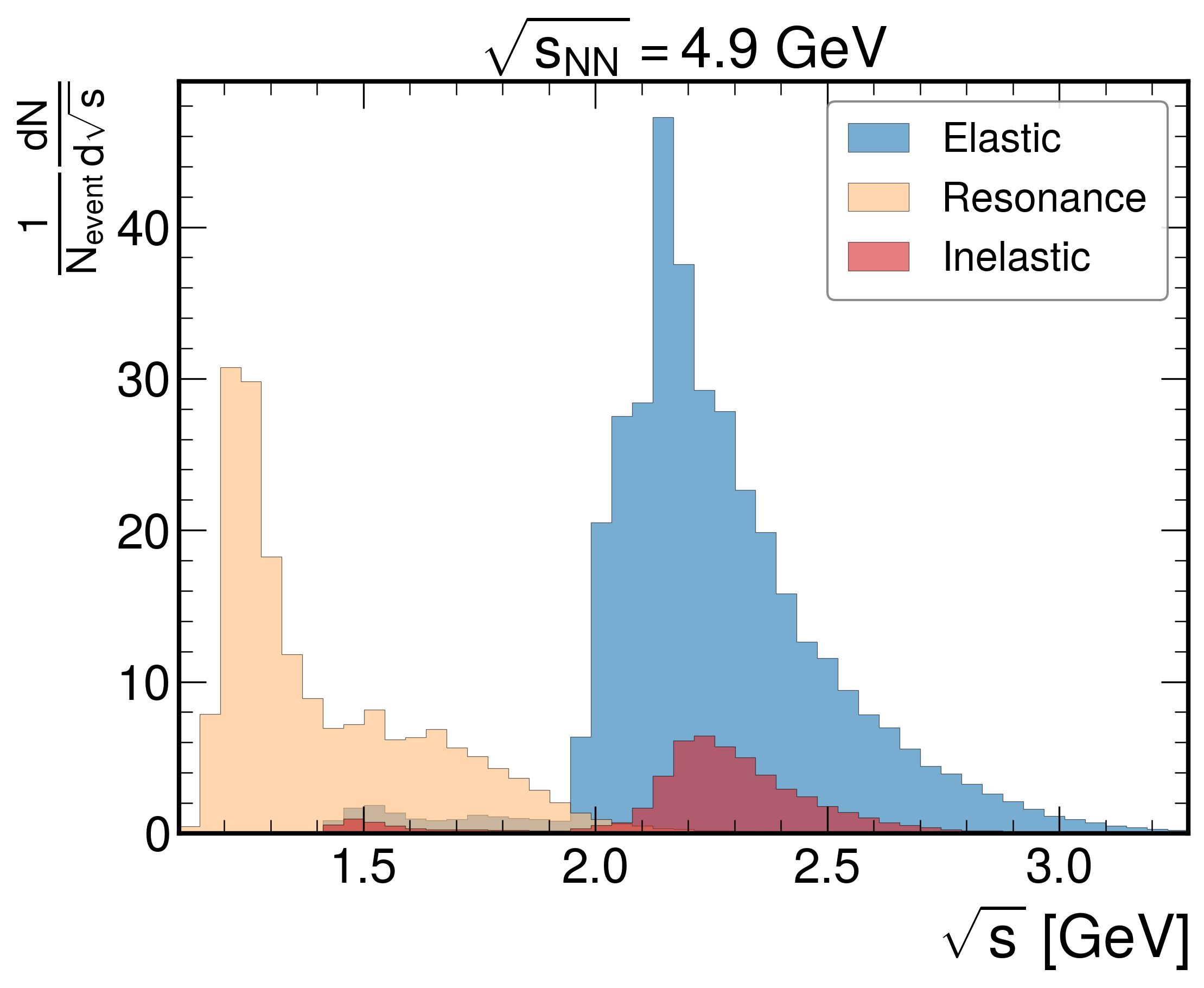}
\caption{Distribution of the proton--baryon center-of-mass energy $\sqrt{s}$ for elastic scattering, resonance formation, and inelastic binary scattering at $\sqrt{s_{NN}}=4.9$~GeV.}
\label{fig:sqrts_proton_4p9}
\end{figure}

Figure~\ref{fig:sqrts_proton_4p9} shows the distribution of the proton--baryon center-of-mass energy $\sqrt{s}$ for the three dominant collision types at $\sqrt{s_{NN}}=4.9$~GeV. Resonance formation is concentrated at low $\sqrt{s}$, reflecting the prominent low-mass baryon resonances; this, together with the high nucleon density at early times, explains the large early-time rate seen in Figure.~\ref{fig:dNcoll_dt_4p9}. Elastic scattering is shifted toward higher $\sqrt{s}$, while inelastic binary scattering occupies the intermediate range above the inelastic threshold.

\begin{figure}
\centering
\includegraphics[width=0.48\textwidth]{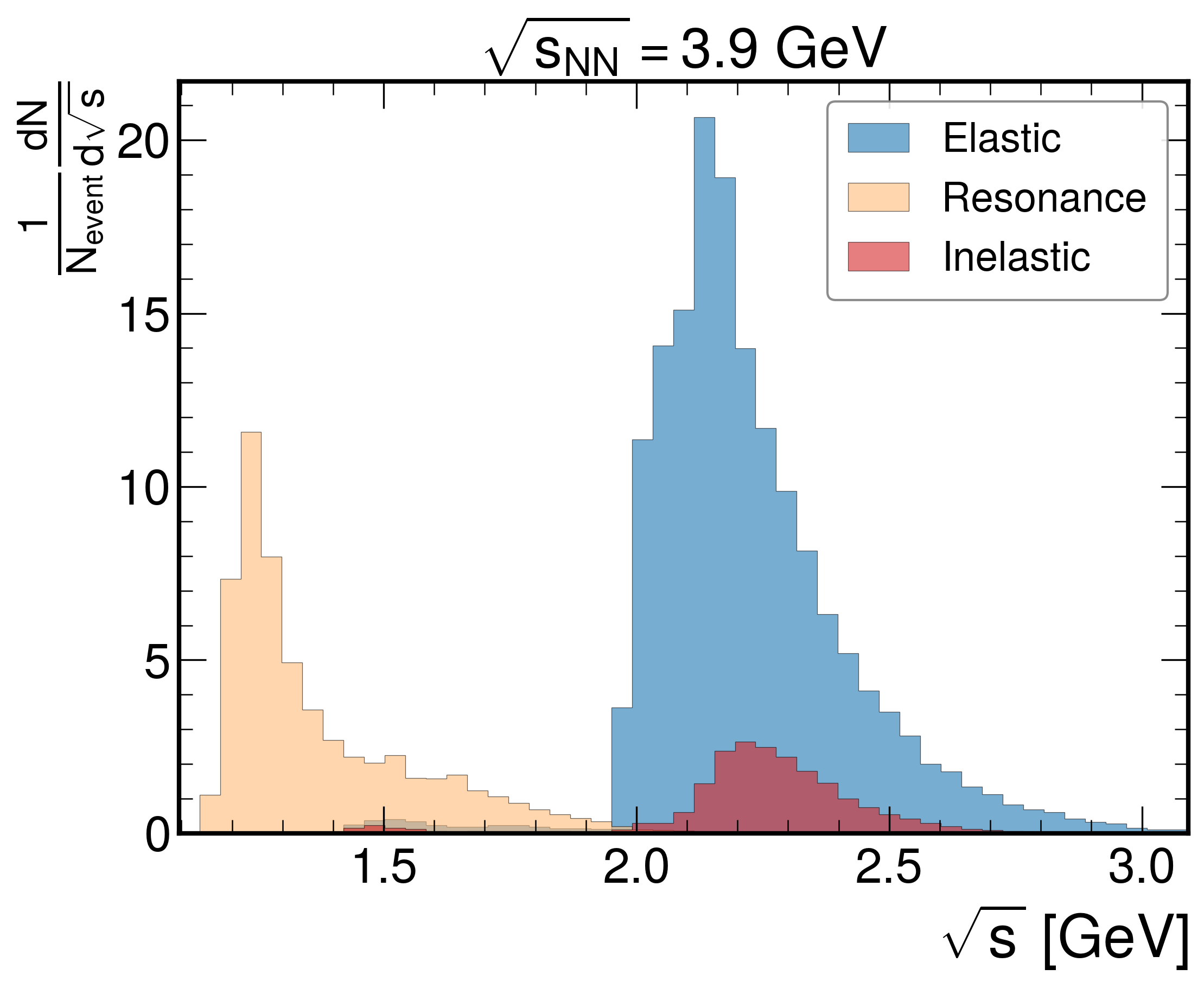}
\caption{Distribution of the proton--baryon center-of-mass energy $\sqrt{s}$ for elastic scattering, resonance formation, and inelastic binary scattering at $\sqrt{s_{NN}}=3.9$~GeV.}
\label{fig:sqrts_proton_3p9}
\end{figure}

The same decomposition at $\sqrt{s_{NN}}=3.9$~GeV is shown in Figure.~\ref{fig:sqrts_proton_3p9}. The dominant collision types remain the same, but the relative weights and the available phase space change with collision energy. The cumulative effect of these small changes in the collision dynamics contributes to the visible difference between the $C_4/C_2$ values at 3.9 and 4.9~GeV seen in Figure.~\ref{fig:CE_C4C2_energy}.

\begin{figure}
\centering
\includegraphics[width=0.48\textwidth]{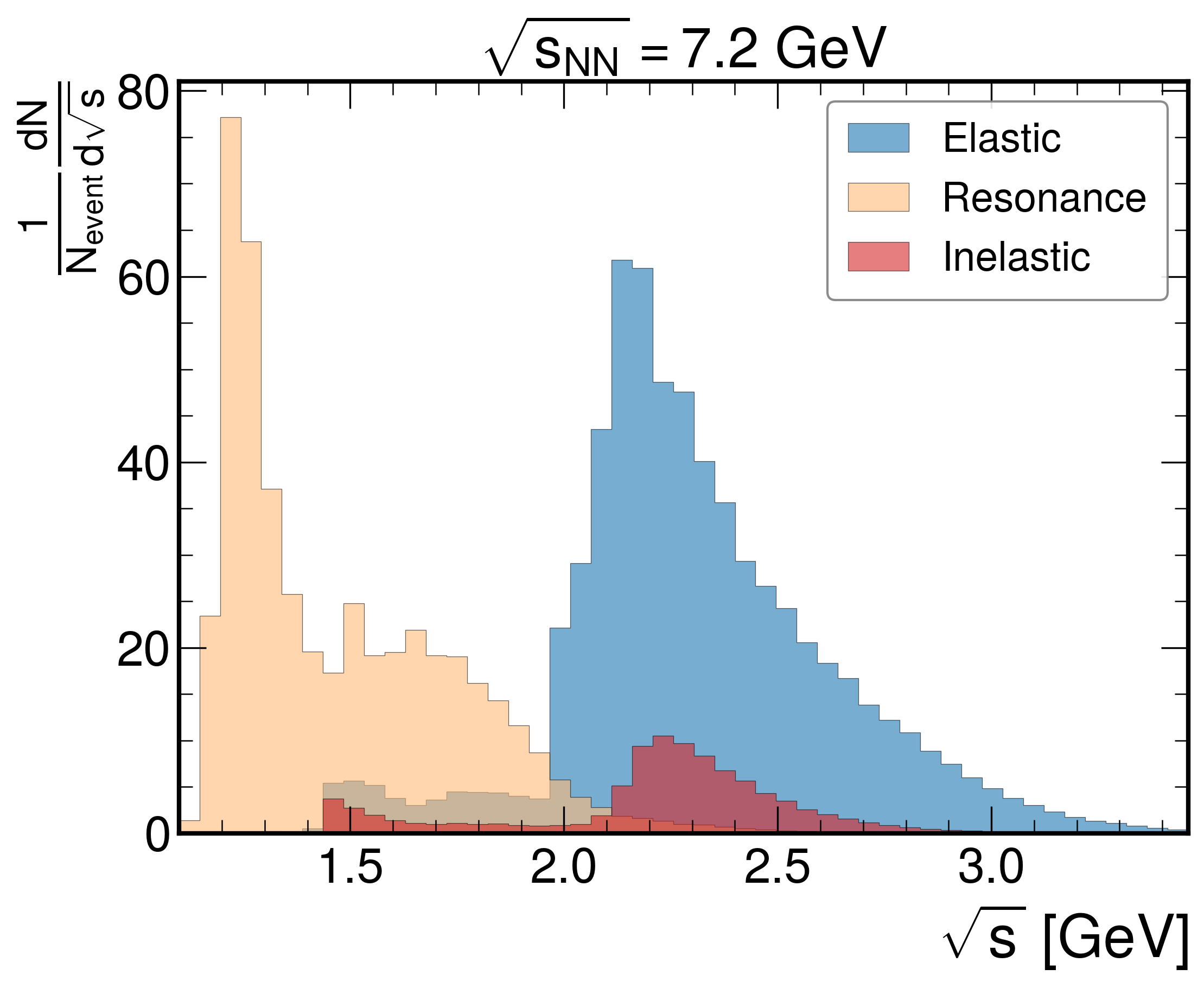}
\caption{Distribution of the proton--baryon center-of-mass energy $\sqrt{s}$ for elastic scattering, resonance formation, and inelastic binary scattering at $\sqrt{s_{NN}}=7.2$~GeV.}
\label{fig:sqrts_proton_7p2}
\end{figure}

For completeness, Figure.~\ref{fig:sqrts_proton_7p2} shows the corresponding proton--baryon center-of-mass energy distribution at $\sqrt{s_{NN}}=7.2$ GeV. Compared with the lower-energy cases, the resonance-formation component exhibits a higher peak in the low-$\sqrt{s}$ region. This suggests that the number of proton--baryon collisions entering the resonance-formation channel increases with collision energy. Such an enhanced resonance contribution provides a microscopic indication of stronger hadronic activity at higher collision energies and helps explain the observed energy dependence of the cumulant evolution.

\section{Conclusion and Outlook}
\label{sec:Conclusion}

In this work, we quantified the impact of hadronic rescattering on
net-proton cumulants by coupling equilibrium cumulants of
baryon-number fluctuations to the SMASH hadronic transport model. The equilibrium cumulants used,  \cite{Fu:2023lcm}, have been computed with the fRG and are based on the QCD phase structure from functional QCD \cite{Fu:2019hdw}. Net-proton multiplicity distributions were reconstructed from the fRG
cumulants via the maximum-entropy principle, converted from the GCE
to the CE through the subensemble acceptance method to enforce exact
net-baryon-number conservation, and propagated event by event through
the hadronic cascade at five BES collision energies.

Our central result is that the effect of hadronic rescattering on the
cumulant ratio $C_4/C_2$ is strongly energy-dependent. It is most
pronounced at the location of the peak at $\sqrt{s_{\rm NN}} = 4.9$~GeV, where the final-state
value in the CE is reduced by approximately $20\%$ relative to its
freeze-out value, while the modifications at the other collision
energies remain mild. At $4.9$~GeV, the suppression occurs
predominantly during the early, dense stage of the cascade and is
followed by a partial recovery at late times, demonstrating that the
hadronic evolution does not act as a simple monotonic damping of
fluctuations.
Since it has been shown in Ref.\ \cite{Fu:2023lcm} that the location of the CEP is encoded in the peak height of $C_4/C_2$, this suppression will play an important role in the identification of the CEP in heavy-ion collisions.

The non-monotonic collision-energy dependence of
$C_4/C_2$ inherited from the fRG input---associated with the assumed
critical endpoint---survives the hadronic stage in both the GCE and
CE frameworks. Rescattering therefore modifies the magnitude of the
signal but does not wash out its qualitative structure. The comparison
between the GCE and CE results further confirms that exact
conservation laws are indispensable for a quantitative description at
the low energies considered here. Our collision-rate analysis
identifies resonance formation at low proton--baryon center-of-mass
energies as the dominant microscopic channel driving these
modifications.

Hadronic afterburner effects on fluctuation observables have been
investigated before. In Ref.~\cite{Hammelmann:2023aza},
maximum-entropy distributions constructed from critical-mode
parametrizations were propagated through a hadronic medium, and
Ref.~\cite{Chen:2024zry} studied the influence of hadronic
rescatterings on net-baryon fluctuations within a transport approach.
The present work differs from these studies in three aspects. First,
the initial cumulants are taken from a direct fRG calculation at finite $\mu_B$, rather than from model parametrizations of
critical dynamics, so that the non-monotonic energy dependence of
$C_4/C_2$ is inherited from a microscopic QCD input. Second, we
implement exact net-baryon-number conservation through the
subensemble acceptance method, which is essential at the low
collision energies considered here. Third, the event-by-event
proton--antiproton reconstruction allows us to track not only the
final-state cumulants but also the microscopic collision dynamics
(rates and $\sqrt{s}$ distributions per channel) responsible for their
modification. Our findings are consistent with the general conclusion
of Refs.~\cite{Hammelmann:2023aza,Chen:2024zry} that the hadronic stage acts as a non-negligible
background, and quantify this effect for fRG-based inputs across the
BES energy range for the first time.

Several systematic uncertainties remain and will be addressed in future work.

\emph{Baryon-to-proton mapping.} We employed the
approximation $C^B_n \simeq C^p_n$ to identify net-baryon with
net-proton cumulants. At the lowest collision energies studied here, the proton fraction among all baryons decreases and contributions from hyperons and baryonic resonances become non-negligible, so this identification introduces a systematic uncertainty. A refined treatment, for instance within a hadron-resonance-gas-based projection of the fRG susceptibilities onto the proton sector, including feed-down and resonance-decay contributions, would provide a quantitative error estimate and is planned as a next step.

\emph{Acceptance fraction in the SAM.} Throughout this work, the
acceptance fraction $\alpha$ was fixed phenomenologically via the
results of Ref.~\cite{Fu:2023lcm}. Since the CE cumulants depend sensitively on
$\alpha$, a dedicated study of the propagation of its uncertainty to
the final-state suppression of $C_4/C_2$ is required. A self-consistent
determination of $\alpha$ directly from the STAR rapidity and
transverse-momentum acceptance, together with a systematic scan in
$\alpha$, is left to a forthcoming analysis.

\emph{Realistic collision geometry.} In the present setup, all particles are sampled within a uniform
sphere and enter the hadronic cascade at a common freeze-out time.
In a realistic hydrodynamic evolution, however, the fireball
hadronizes gradually from its surface inward: during most of the
evolution, only a shell near the surface at $R \sim 6$~fm crosses
the freeze-out hypersurface, and only in the late, dilute tail stage
does hadronization occur simultaneously over the whole volume. As a
consequence, fluid cells hadronized at different times experience
different durations of hadronic rescattering, which is expected to
modify quantitatively the cumulant evolution found here.
The extracted modification of the cumulants should therefore be
regarded as an estimate at fixed average density. Coupling the
present framework to a realistic fluctuating initial state, e.g.\ a
hybrid (viscous) hydrodynamic evolution followed by the SMASH
afterburner, will allow an event-by-event study with realistic
centrality and acceptance definitions.
It has been shown in \cite{Fu:2023lcm} that the location of the peak of the fluctuations observables is rather  insensitive to the location of the CEP but the height and shape is. Future studies will extend this analysis and also systematically assess how the predicted scaling of fluctuation observables and their peak structures depend on the assumed location of the critical endpoint, see e.g.~\cite{Du:2024wjm}.

\begin{acknowledgments}

The authors thank Jiaxing Zhao for useful discussions and comments. This work is supported by the China Scholarship Council (CSC) under Grant No. 202406770085. This work is also supported by the National Natural Science Foundation of China under Grant No.\ 12075098, No.\ 12435009, by the Outstanding Leading Talent Team Program of Central China Normal University (XJ2026000302), DFG Collaborative Research Centre ``SFB 1225 (ISOQUANT)''. The numerical calculation have been performed on the GPU cluster in the Nuclear Science Computing Center (NSC3) at CCNU. S. Yin is supported by the Alexander von Humboldt foundation.
\end{acknowledgments}

\bibliography{ref}

\end{document}